\documentclass[useAMS,usenatbib]{mn2e}
\usepackage{graphicx}
\usepackage{amsmath}
\newcommand{\ltaraw}{$\; \buildrel < \over \sim \;$}
\newcommand{\lta}{\lower.5ex\hbox{\ltaraw}}
\newcommand{\gtaraw}{$\; \buildrel > \over \sim \;$}
\newcommand{\gta}{\lower.5ex\hbox{\gtaraw}}

\loadboldmathitalic   \title[Is the FB model concordant?]{Fractal Bubble Cosmology: 
A concordant cosmological model? \footnotemark[1]} 
 \author[Kwan et al.]{Juliana Kwan$^{1}$, Matthew J. Francis$^{1,2}$ \&  Geraint F.
  Lewis$^{1}$ \\
  $^{1}$Sydney Institute for Astronomy, School of Physics, A28,
  University of Sydney, NSW 2006, Australia\\
  $^2$SISSA, International School of Advanced Studies, via Beirut 2-4, 34014, Trieste, Italy\\
} 
\date{\today}
\begin{document}
\maketitle
\label{firstpage}
\begin{abstract}
The Fractal Bubble model has been proposed as a viable cosmology that
does not require dark energy to account for cosmic acceleration, but
rather attributes its observational signature to the formation of
structure. In this paper it is demonstrated that, in contrast to
previous findings, this model is not a good fit to cosmological
supernovae data; there is significant tension in the best fit
parameters obtained from different samples, whereas $\Lambda$CDM is
able to fit all datasets consistently.  Furthermore, the concordance
between galaxy clustering scales and data from the cosmic microwave
background is not achieved with the most recent supernova
compilations. The validity of the FB formalism as a sound cosmological
model is further challenged as it is shown that previous studies of
this model achieve concordance by requiring a value for the present
day Hubble constant that is derived from supernovae data containing an 
arbitrary distance normalisation.

 \end{abstract}

\long\def\symbolfootnote[#1]#2{\begingroup%
  \def\thefootnote{\fnsymbol{footnote}}\footnotetext[#1]{#2}\endgroup} 

\def\newblock{\hskip .11em plus .33em minus .07em}
\section{Introduction}     \label{intro}     \symbolfootnote[1]{Research
  undertaken as part of the Commonwealth Cosmology Initiative (CCI:
  www.thecci.org), an international collaboration supported by the
  Australian Research Council} The discovery of cosmic
  acceleration,~\citep{Riess98, Perlmutter99}, and its subsequent
  explanation as dark energy has resulted in the proposal of many
  alternate models to the current cosmological paradigm $\Lambda$ Cold
  Dark Matter ($\Lambda$CDM) as the nature of dark energy remains
  elusive. There are three main alternatives to dark energy: either
  General Relativity is not the correct description of gravity on
  large scales, introduce a higher dimensional theory, or the metric
  that describes the universe is not required to be maximally
  symmetric~\citep[see][for a review]{Durrer08}. The Fractal
  Bubble (FB) model~\citep{Wiltshire07a,Wiltshire07b} falls into the
  last category, as dark energy is replaced by `quasi-local
  gravitational energy' arising from changes in the curvature of
  spacetime due to an inhomogeneous distribution of
  matter. In~\cite{Leith08}, it is claimed that the FB model achieves
  a concordant set of cosmological parameters using supernova data
  from~\cite{Riess07}, the angular scale of the sound horizon from
  WMAP1~\citep{Bennett03} and the comoving spatial separation of the
  correlation function in SDSS~\citep{Eisenstein05}.

Although the vast majority of inhomogeneous cosmological models have
been shown to be inconsistent with observations or rely on a
particular choice of
coordinates~\citep{Ishibashi06,Ziblin08,Caldwell08}, the FB model
remains unchallenged. In this paper, we examine the ability of the FB
model to produce a concordant set of cosmological parameters. In
particular, we address the claim of~\cite{Leith08} that their two
parameter fits for the distance modulus given by the FB model using
the supernova type Ia (SNe Ia) sample of~\cite{Riess07} supports a
concordance of observational evidence for the FB model. Additionally,
we demonstrate a number of discrepancies including the inability of
the FB model to consistently describe both the~\cite{Riess07} sample
and the Union supernova compilation~\citep{Kowalski08} or the
Constitution set~\citep{Hicken09} despite none of these SNe Ia samples
being in tension. \cite{Leith08} also omit to calibrate 
the~\cite{Riess07} SNe Ia data to a distance scale when fitting for the
Hubble constant, H$_0$.

\begin{figure} 
\begin{center}
\includegraphics[scale = 0.5,angle=270]{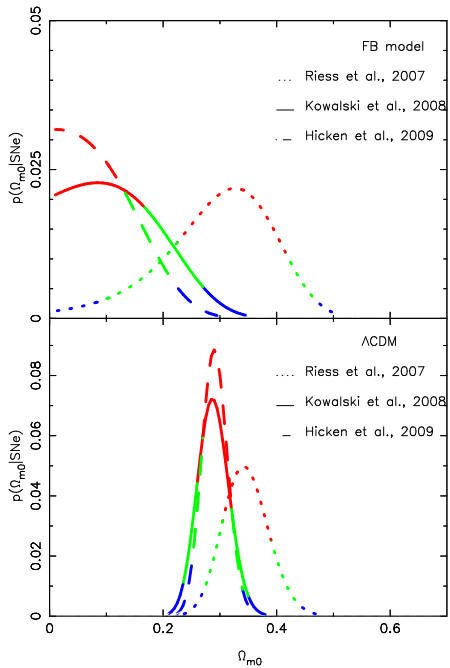}
\caption[]{\emph{Top}: Marginalised posterior distributions for the FB
model using the Union compilation~\citep{Kowalski08} in solid lines,
the Gold SNe Ia sample from~\cite{Riess07} in dotted lines, and the
Constitution set~\citep{Hicken09} in dashed lines.  The colours
correspond to the confidence limits, red is 1$\sigma$, green is
2$\sigma$ and blue is 3$\sigma$. \emph{Bottom}: As
above except using $\Lambda$CDM.
\label{fig:posterior_margH0}} 
\end{center}
\end{figure} 

\section{Background}
The FB model~\citep{Wiltshire07a,Wiltshire07b} is a two scale model
with a local r\'{e}gime for over or underdense regions and another for
the volume average, which utilises the Buchert averaging
scheme~\citep{Buchert00}. We use an overbar to indicate volume
averaged quantities as opposed to `dressed' parameters measured by
observers in galaxies, termed `walls' in the FB model. The volume
averaged scale factor, $\bar{a}$, and the void fraction, $f_v$,
defined as the total volume in void regions, evolve according to the
Buchert equations as follows:
\begin{eqnarray}
\frac{\dot{\bar{a}}^2}{\bar{a}^2} + \frac{\dot{f_v}^2}{9f_v(1-f_v)} - \frac{\alpha^2 f_v^{1/3}}{\bar{a}^2} = \frac{8\pi G}{3}\bar{\rho}_0 \frac{\bar{a}_0^3}{\bar{a}^3}, \\ \label{eqn:buchert}
\ddot{f_v} + \frac{\dot{f_v}^2(2f_v-1)}{2f_v(1-f_v)} + 3\frac{\dot{\bar{a}}}{\bar{a}}\dot{f_v} - \frac{3\alpha^2 f_v^{1/3}(1-f_v)}{2\bar{a}^a} = 0,
\end{eqnarray}
where $\alpha$ is the normalisation constant of the curvature energy.
Equation 1 is equivalent to writing: \mbox{$\bar{H}^2 -
\bar{\Omega}_q\bar{H}^2 - \bar{\Omega}_k\bar{H}^2 =
\bar{\Omega}_m\bar{H}^2$}, since the volume averaged Hubble constant
is defined as $\bar{H} = \dot{\bar{a}}/\bar{a}$ and the
normalised volume averaged energy densities are given by:
\begin{align}
\bar{\Omega}_q = -\frac{\dot{f_v}^2}{9f_v(1-f_v)\bar{H}^2}  &  \quad {\rm (backreaction)}, \\
\bar{\Omega}_k = \frac{\alpha^2 f_v^{1/3}}{\bar{a}^2\bar{H}^2}  &  \quad {\rm (curvature)}, \\
\bar{\Omega}_m = \frac{8\pi G}{3}\bar{\rho}_0 \frac{\bar{a}_0^3}{\bar{a}^3\bar{H}^2} &  \quad {\rm (matter)}.
\end{align}  
Information about the distribution of matter is encoded in the void
fraction, $f_{v}$, which is directly proportional to the dressed
normalised matter density, $\Omega_{m,0}$, as $\Omega_{m,0} \approx
\frac{1}{2}(1-f_{v,0})(2+f_{v,0})$~\citep{Wiltshire07b}. Although a
backreaction term occurs in the above equations, this is not the
principal mechanism that creates cosmic acceleration, but rather this
is achieved through `quasi-local' gravitational energy quantified by
the curvature energy term. Such gravitational energy can not be
localised in the stress energy tensor and in the FB model gives rise
to an apparent cosmic acceleration as measurements are distorted when
signals pass through regions of varying curvature and it cannot be
assumed that the clocks of all observers are synchronous independent
of location. In the absence of homogeneity, volume averaged quantities
must be transformed back to what observers in overdensities like
ourselves would measure. The FB model, chooses to foliate spacetime
such that there is a lapse function, defined as $\bar{\gamma}(\tau)
\equiv dt/d\tau$, where $\tau$ is the proper time of an observer in a
wall or a void and $t$ is the volume averaged or cosmic time, which
quantifies the difference between clocks of two different observers
depending on their location. If the lapse function in walls is
significant, then the FB model predicts that a spurious cosmic
acceleration will be detected by failing to account for an
inhomogeneous matter distribution through the assumption of a
Friedmann-Lema\^{i}tre-Robertson-Walker (FLRW) background and a
recalibration of observations for the difference in clocks between
walls and the volume average is necessary. This difference may be as
large as $\bar{\gamma} \approx 1.38$ to achieve the level of
concordance seen in~\cite{Leith08}.

Unlike previous approaches with inhomogeneous cosmologies, much of the
physics behind the FB model is essentially alien to the standard model
of modern cosmology. An extension to the equivalence
principle~\citep{Wiltshire08} is required to explain the anomalously
large value of $\bar{\gamma}$ necessary to produce the best fit
parameters quoted in~\cite{Leith08}
and~\cite{Wiltshire07a}. Furthermore, neither the Newtonian limit or
Birkhoff's theorem is relevant in this cosmology. Leaving aside these
conceptual issues, we have instead focused on establishing the
compatibility of currently available observational data with the FB
model.

\section{Tension with supernova data}
\subsection{Constraints from new data sets}
The Union supernova compilation~\citep{Kowalski08} is primarily
composed of SNe Ia catalogs from literature that have been reanalysed
in an uniform manner to reduce systematics, with some further cuts
imposed to exclude data of insufficient quality. The compilation
contains a total of 307 SNe Ia, of which 27 are drawn
from~\cite{Riess07} after the final cut and a further eight are
derived from new low-redshift observations. Its constituent catalogues
have also been shown to be consistent with one another under this new
analysis [see Figure 9 of~\cite{Kowalski08}]. More recently, the
entire Union compilation has been amalgamated with an additional
sample of 90 new CfA3 SNe Ia into the Constitution set, increasing the
amount of low redshift data by a factor of 2.6~\citep{Hicken09}. Like
the Union compilations, the Constitution set is uniformly reduced
through the use of a single light curve fitter for all SNe
Ia. Clearly, both the Union compilation and the Gold sample
of~\cite{Riess07} are entirely consistent with the Constitution set.

We test the ability of the FB model to constrain cosmology with the
new Union and Constitution SNe Ia data sets, as well as the Gold
sample from~\cite{Riess07}. This last sample was analysed
by~\cite{Leith08}, but this study suffered from a number of problems
as, detailed in Section 3.2, which we rectify in this work. The best
fit cosmological parameters are found from calculating the reduced chi
squared, $\chi^2$, based on the distance modulus given by $\mu =
5\log_{10}(d_L) + 25$ for $\Lambda$CDM and FB model to all three
samples of SNe Ia. The luminosity distance $d_L$, is defined as:
\begin{equation}
d_L = \left\{
\begin{array}{cc}
(1+z)a_0\int^{t_0}_t \frac{dt}{a(t)}   &  {\rm (\Lambda CDM)}, \\
\frac{\bar{\gamma}}{\bar{\gamma}_0}\bar{a}_0(1+z){(1-f_v)}^{1/3}\int^{t_0}_t \frac{dt}{\bar{\gamma}{(1-f_v)}^{1/3}\bar{a}}  &  {\rm (FB)}. \\
\end{array} \right.
\nonumber
\end{equation}
There are two free parameters of interest in the FB model that are
used to calculate the distance modulus; the cosmology is fully
constrained after specifying the initial void fraction $f_{vi}$, the
dressed Hubble constant today, H$_0$, and the initial ratio of the
expansion rates as measured in a wall and a void, $h_{ri}$; the latter
remains fixed for all fits. We have used the same priors
as~\cite{Leith08} on f$_{vi}$ and h$_{ri}$, namely $10^{-5} \: < \:$
f$_{vi} \: < \: 10^{-2}$ (and hence 0.01 $\leq \Omega_{m,0} < 0.95$)
and h$_{ri} = 0.99999$ at last scattering, with 50 $ \leq H_{0} \leq
75$ kms$^{-1}$Mpc$^{-1}$. The $\Lambda$CDM priors are the same on
H$_0$ and cover a similar range on $\Omega_{m,0}$: $0 \leq
\Omega_{m,0} \leq 1$. The value of H$_0$ is treated as a nuisance
parameter, which we marginalise over, since it cannot be generally
assumed that the absolute magnitudes of the SNe Ia are known. This
results is a single parameter fit for $\Omega_{m,0}$, which is
directly related to $f_v$.

\begin{figure} 
\begin{center}
\includegraphics[scale = 0.3,angle=270]{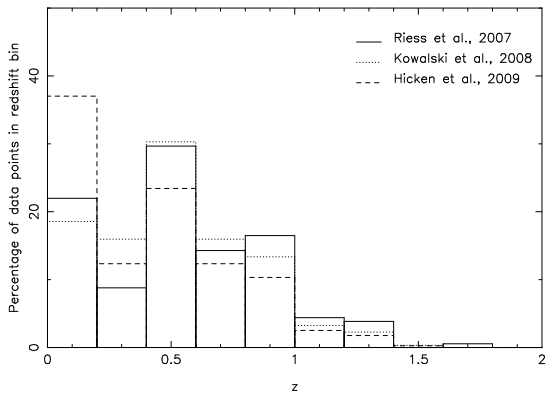}
\caption[]{Percentage of SNe Ia from Gold (solid), Union (dotted) and
Constitution (dashed) samples falling into redshift bins of width
$\Delta$z = 0.2. There are 397 SNe Ia in the Constitution set, 307 SNe
Ia in the Union compilation and 182 SNe Ia in the Gold data set
of~\cite{Riess07}.
\label{fig:redshift_dist}} 
\end{center}
\end{figure} 

The $\chi^2$ statistic is converted to a likelihood, $\mathcal{L}$,
via the relationship, \mbox{$\mathcal{L} \propto
\exp({-\chi^2/2})$}. Bayesian inference then gives the resultant
posterior distributions as presented in
Figure~\ref{fig:posterior_margH0} for both cosmological models, with
the best fit values shown in Table~\ref{tab:bestfit_params}. For
$\Lambda$CDM, all three distributions give a consistent set of
cosmological parameters that are in agreement, with the 1$\sigma$
confidence regions significantly overlapping. This does not occur for
the FB model; the 1$\sigma$ limits derived for each SNe Ia sample do
not all coincide. We must consider the 2$\sigma$ confidence region
before we can find a value of $\Omega_{m,0}$ that agrees with between
the Union compilation or the Constitution set and the Gold sample
of~\cite{Riess07}. In fact, the best fit value for the Gold sample
from~\cite{Riess07} is ruled out at 3$\sigma$ when performing the same
analysis for the Union compilation and is beyond the 3$\sigma$ limit
for the Constitution set. (See Table~\ref{tab:bestfit_params} for the
complete set of best fit parameters and 1$\sigma$ errors).  However,
there is good agreement between the best fit FB parameters for
Constitution set and the Union compilation, which is reassuring; the
reason for this discordance lies with the model rather than an anomaly
in the SNe Ia sample. If this model is to be acknowledged as a viable
alternative to $\Lambda$CDM, then it seems that we must also accept
that there is no consistent observational evidence from SNe Ia for the
FB model or that these two SNe Ia samples are in tension with
the~\cite{Riess07} data set. However,~\cite{Kowalski08} found a high
degree of consistency between the samples used in the construction of
the Union compilation, albeit with mild tension between
the~\cite{Krisciunas04a,Krisciunas04b,Krisciunas01} and~\cite{Hamuy96}
samples and the other SNe Ia data when comparing $d\mu/dz$. Crucially,
the more recent and populous samples break the concordance of
cosmological tests for the FB model; clearly the best fit values for
$\Omega_{m,0}$ as shown in Table~\ref{tab:bestfit_params} do not
overlap with those obtained from galaxy clustering statistics and the
angular scale of the sound horizon found by~\cite{Leith08}, which
require 0.27 $\la \Omega_{m,0} \la $ 0.37. Furthermore, although the
Bayes factor~\citep{Trotta08} between the FB model and $\Lambda$CDM
with the~\cite{Riess07} Gold data set marginally favours the former
[lnB = 0.27,~\cite{Leith08}], the FB model is weakly disfavoured when
using the Union compilation or the Constitution set (lnB = -1.38,
-1.469, respectively) under the same classification scheme on the
strength of evidence. As the treatment of systematics improve and
sample sizes grow, $\Lambda$CDM is increasing being favoured by the
data while the best fit parameters are moving away from those required
by the FB model for concordance. The deterioration of the evidence
towards the FB model in terms of the Bayes factor may be attributed to
the width of the posterior in $\Omega_{m,0}$, which remains
significantly larger than that of the corresponding data set under
$\Lambda$CDM.

\begin{figure}
\begin{center}
\includegraphics[scale = 0.55, angle=270]{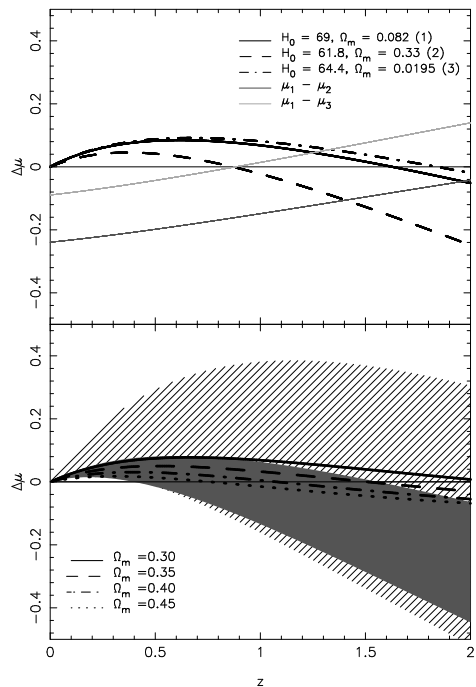}
\caption[]{{\it Left}: Residuals from an empty FLRW universe with same
value of H$_0$ as the FB model with best fit parameters from the Union
compilation (solid), Constitution set (dot-dashed) and
Gold~\cite{Riess07} sample (dashed). The difference in the distance
moduli between the best fit values found by~\cite{Leith08} and those
from the more recent compilations are shown in light grey (Union) and
dark grey (Constitution) {\it Right}: Range of residuals from an empty
FLRW universe with same value of H$_0$ for FB model (solid grey) and
$\Lambda$CDM (hatched black) for 0.1 $ < \; \Omega_{m,0} \; <$
0.5. Any residual from an empty universe using either the FB model or
$\Lambda$CDM, with the above parameters, will fall into the
corresponding shaded area. The curves show the difference between the
distance moduli of the FB model and $\Lambda$CDM for the same value of
H$_0$ for $\Omega_{m,0} = {0.30 \: (solid), 0.35 \: (dashed),
0.40 \: (dot-dashed), 0.45 \: (dotted)}$. There is a preferred range of
parameters, namely 0.35 $\la \Omega_{m,0} \la$ 0.4 for which the
difference between the two models is minimised.
\label{fig:combined_residuals}}
\end{center}
\end{figure}

\subsection{Distance normalisation?}
Figure 2 of~\cite{Leith08} depicts a concordance diagram, represented
in the $\Omega_{m,0}$-H$_0$ plane, for the FB model, with confidence
limits from SNe Ia using the Gold data of~\cite{Riess07}, the proper
distance to the sound horizon seen in SDSS, and the angular scale of
the sound horizon at decoupling. However, a number of issues regarding
the analysis of these observational constraints remain unaddressed by
the authors, which, when corrected, falsify the concordance. The most
serious of these is that the SNe Ia data used to produce this diagram
contains an arbitrary value of H$_0$ without the appropriate correction
to the distance moduli. Thus, it is misleading to present confidence
limits in Figure 2 of~\cite{Leith08} that depend on H$_0$, since these
do not represent its true value. Instead, the concordance diagram
in~\cite{Leith08} should be shown with either H$_0$ as a nuisance
parameter, which is marginalised or with the recommended calibration
of~\cite{Riess07}.

For the Gold sample,~\cite{Riess07} suggest using a systematic
subtraction of 0.32 mag\footnote{Magnitude calibration and data can be
obtained online at {\tt
braeburn.pha.jhu.edu/$\sim$ariess/R06/sn\_sample}} to use the same
Cepheid scale as~\cite{Riess05} if the value of H$_0$ is of interest
and this is unnecessary for fits to dynamic quantities only. An
arbitrarily chosen value of H$_0$ is inserted into the data, since the
absolute magnitude of a SNe Ia is not known until an appropriate
distance scale, such as that obtained from Cepheid luminosities, is
applied. This correction to the distance moduli has not been taken
into account by~\cite{Leith08}, who claim to follow the Cepheid
calibration of~\cite{Sandage06} but no corrections have been applied
at all. Indeed, when performing our fits to the~\cite{Riess07} Gold
sample, we are able to reproduce the same best fit value of H$_0$ as
quoted in~\cite{Leith08} (within 0.1 error) without accounting for the
arbitrary distance normalisation at all. The authors, however, have
interpreted the value of H$_0$ in the~\cite{Riess07} sample as a
physical parameter that constrains cosmology, rather than as a value
chosen by~\cite{Riess07}. Thus the fit to H$_0$ in Figure 2
of~\cite{Leith08} is meaningless without calibrating the data. It is
plausible that the arbitrary normalisation chosen by~\cite{Riess07}
happens to coincide with the Sandage distance scale, but no
justification is given for the choice of calibration
in~\cite{Leith08}. Interestingly, using the Cepheid calibration
recommended by~\cite{Riess07} gives a best fit value of H$_0 = 71.6$
kms$^{-1}$Mpc$^{-1}$ with $\Omega_{m,0}$ unchanged, since changing
H$_0$ only shifts the scale of the Hubble diagram. This no longer
produces a concordant set of cosmological parameters; in fact it is
stated in~\cite{Leith08} that for any value of H$_0$ greater than 70
kms$^{-1}$Mpc$^{-1}$, the observational constraints from the proper
distance to the sound horizon and its angular scale at decoupling
would no longer agree.

\section{A Coincidence Problem for the FB model}
To understand why the cosmology of the FB model changes dramatically
for the Union compilation and the Constitution set while $\Lambda$CDM
remains consistent, we have analysed the redshift distributions of
these two SNe Ia catalogues.  Figure~\ref{fig:redshift_dist} shows
that the Gold sample of~\cite{Riess07} contains a slightly higher
percentage of high redshift SNe Ia than in the other samples but there
are significantly more low redshift samples in the latter,
particularly in the 0.0  $ \leq z  < $0.2 range for the
Constitution set and 0.35 $\la  z < $  0.4 range for the
Union compilation. Furthermore, it should be unsurprising that the FB
model is able to fit both the Constitution set and the Union
compilation within 1$\sigma$ error; Figure~\ref{fig:redshift_dist}
also shows that the redshift distribution of these two samples are
very similar, especially in the high-$z$ r\'{e}gime where the
Constitution set is dominated by data taken from the Union
compilation.

The reason behind the shift in parameter space is made apparent by
considering the range over which the newer compilations contain more
data. This is coincident with the region in which the three different
best fits for the FB model diverge the most
(Figure~\ref{fig:combined_residuals}; left). The residuals have been
taken relative to an empty universe with $\Omega_{m,0} = 0, \;
\Omega_{\Lambda,0} = 0, \; \Omega_{k,0} = 1$, and the same H$_0$ to
remove the uncertainty in its value. For the Union and Constitution
sets, the FB model requires a much greater amount of acceleration to
fit the data and indeed the period of apparent acceleration occurs at
$z \approx \; 1.65$ and $z \approx \; 1.85$ respectively. As the
cosmological constraints become progressively tighter in the more
recent SNe Ia samples, the FB model is increasingly pushed to higher
values of curvature energy. Indeed, the best fit parameters from the
Union and Constitution sets imply that the universe is almost entirely
composed of voids with an almost negligible amount of collapsed
structure. As remarked in the previous section, the FB model has a
flatter, wider posterior than $\Lambda$CDM, which implies that a
greater range of the parameter space can fit the data. This can be
attributed to the limited range of behaviour that the FB model can
exhibit in terms of changes to the distance modulus when $f_{vi}$ is
altered. This is demonstrated in the difference between the range that
is spanned by the residuals for the FB model (solid grey) and
$\Lambda$CDM (hatched black) for 0.1 $ < \; \Omega_{m,0} \; <$ 0.5
(Figure~\ref{fig:combined_residuals}, right). Although the prior on
$f_{vi}$ spans $\approx$ 3 orders of magnitude, the residuals for the
FB model have a substantially truncated r\'{e}gime of acceleration in
comparison to $\Lambda$CDM. This partly accounts for the discordance
seen in the confidence limits for this model that is not replicated in
$\Lambda$CDM. The best fit parameters of the FB model are extremely
sensitive to small changes in the SNe Ia data as it needs to
compensate for these by a large variation in $f_{vi}$ when fitted to an
another redshift distribution with a different amount of error on
each SNe Ia. In addition, there is a special set of values for $f_{vi}$
which will mimic $\Lambda$CDM parameters well, that is the dressed
matter density, $\Omega_{m,0}$, in the FB model is a similar value to
$\Omega_{m,0}$ derived from $\Lambda$CDM.

In the right panel of Figure~\ref{fig:combined_residuals}, we have
also considered the difference in distance moduli between $\Lambda$CDM
and FB models (each with the same value of H$_0$). This is minimised
when 0.35 $\la \Omega_{m,0} \la$ 0.4 for both models. While it is
strictly not necessary for the FB model to predict the same value of
$\Omega_{m,0}$ as $\Lambda$CDM, since its value is not directly
observable but inferred from theory, it does so because the SNe Ia
sample of~\cite{Riess07} gives a best fit value of $\Omega_{m,0} =
0.34$ for $\Lambda$CDM and this happens to fall within this parameter
range. Since the best fit parameters for both the Union and
Constitution set are $\Omega_{m,0}= 0.29$ and $\Omega_{\Lambda,0} =
0.71$ for $\Lambda$CDM, fitting the FB model to this dataset cannot
produce a similar value for $\Omega_{m,0}$. In addition, changes in
the data that are insignificant for CDM cosmologies will have
sufficient leverage to skew the posterior of FB models as a result of
the similarity of the solutions when $f_{vi}$ is varied. This
behaviour can be attributed to the existence of a tracker
solution~\citep{Wiltshire07b} such that all solutions converge at high
redshift, which inhibits the range of possible behaviour that the FB
model can exhibit.

\begin{table}
\begin{center}
\begin{minipage}[t]{\linewidth}
\caption{Best fit parameters for $\Lambda$CDM and Fractal Bubble model
to SNe Ia data. Distance normalisation is arbitrary for the Union
compilation and the Constitution set. Best fit values of
$\Omega_{m,0}$ and H$_0$ are subject to an additional error of $\pm 5
\times 10^{-3}$ and $\pm 0.1$ respectively due to grid
spacing. \label{tab:bestfit_params}}
\begin{tabular}{lllc}
\hline\hline
Model        & SNe Ia sample                       & $\Omega_{m,0}$ & H$_{0}$ (km/s/Mpc)  \\
\hline
FB           & \cite{Riess07}\footnotemark[1]      & $0.33^{+0.07}_{-0.11}$   & 61.8\\ [1ex]
$\Lambda$CDM & \cite{Riess07}\footnotemark[1]      & $0.34^{+0.045}_{-0.035}$ & 62.8\\ [1ex]
FB           & \cite{Riess07}\footnotemark[2]      & $0.33^{+0.07}_{-0.11}$   & 71.6\\ [1ex]
$\Lambda$CDM & \cite{Riess07}\footnotemark[2]      & $0.34^{+0.045}_{-0.035}$ & 72.8\\ [1ex]
FB           & \cite{Kowalski08}                   & $0.08^{+0.08}_{-0.08}$   & ---\\ [1ex]
$\Lambda$CDM & \cite{Kowalski08}                   &$0.29^{+0.03}_{-0.025}$   & ---\\ [1ex]
FB           & \cite{Hicken09}                     & $0.02^{+0.11}_{-0.02}$   & ---\\ [1ex]
$\Lambda$CDM & \cite{Hicken09}                     & $0.29^{+0.025}_{-0.02}$  & ---\\ 
\hline
\end{tabular}
\end{minipage}\hfill
\begin{minipage}[t]{\linewidth}
$^1$Without calibration \\
$^2$With calibration
\end{minipage}
\end{center}
\end{table}

\section{Conclusions}
One of the more problematic features of the FB model is its failure to
provide any predictions on cosmological parameters that are directly
observable. Although it produces a value of H$_0$ that is consistent
with the HST Key Project~\citep{Freedman01} and observations from WMAP
when fitted to the~\cite{Riess07} SNe Ia sample with the appropriate
Cepheid calibration, this value is not concordant with that obtained
from the proper distance to the sound horizon observed in
SDSS~\citep{Eisenstein05} or the angular scale of the sound horizon
observed in WMAP1~\citep{Bennett03}. Furthermore, the predictions that
are made require further investigation of the model at a fundamental
level; the constrains from differing SNe Ia sample are
discordant. Regardless of the calibration chosen for the Gold sample
of~\cite{Riess07}, there is no value of H$_0$ that can produce a
consistent cosmology within the FB framework under the currently
available SNe Ia data.

The most appealing feature of the FB model is to offer a mechanism for
replacing dark energy with an inhomogeneous matter distribution and
yet it lacks any formalism by which structure formation can
occur. Despite the void fraction being a fundamental parameter of the
FB model, it has not been made transparent in
either~\cite{Wiltshire07a, Wiltshire07b} or~\cite{Leith08} how this is
manifested in measurements of clustering statistics or the growth of
structure. Indeed, the observational evidence for the FB model
presented by~\cite{Leith08} are all derived from geometrical tests and
it is not clear what behaviour it would exhibit under more dynamically
oriented probes such as large scale structure surveys or constraints
from weak lensing. Although it is claimed in~\cite{Leith08} that the
`concordance' values of $\Omega_{m,0}$ derived from the FB model are
more compatible with those obtained from X-ray measurements of cluster
counts~\citep{Yepes07}, it is in fact impossible to state what the FB
model would predict without a mechanism for structure formation,
since such estimates of $\Omega_{m,0}$ are model dependent. Until the
FB model can provide more plausible observational evidence, it is
difficult to envisage this model as a serious competitor to
$\Lambda$CDM when so many questions remain.

\section*{Acknowledgments}
The authors wish to acknowledge support from ARC Discover Project
DP0665574 and would like to thank Eric Linder for his comments and
suggestions.

\end{document}